\documentclass{aa}
\usepackage{graphics,epsfig}

\newcommand{\hcm}[1]{$\cdot 10^{#1}$ cm$^{-2}$}
\newcommand{\ohcm}[1]{$10^{#1}$ cm$^{-2}$}

\newcommand{\nh}{\hbox{N$_{\rm H}$}}

\newcommand{\xmc}{0536.9$-$6913} 
\newcommand{\xmd}{0537.1$-$6913} 
\newcommand{\xme}{0537.2$-$6913} 
\newcommand{\xmf}{0537.3$-$6915} 
\newcommand{\xmh}{0537.5$-$6919} 
\newcommand{\xmi}{0538.0$-$6916} 

\begin{document}
 

\title{AGN in the XMM-Newton first-light image as probes\\
for the interstellar medium in the LMC$^*$}
 
\author{F.~Haberl\inst{1} \and
        K.~Dennerl\inst{1} \and
        M.D.~Filipovi\'c\inst{1,2,3} \and
        B.~Aschenbach\inst{1} \and
        W.~Pietsch\inst{1} \and
        J.~Tr\"umper\inst{1}}

\titlerunning{AGN in the XMM-Newton first-light image}
\authorrunning{Haberl et al.}
 
\offprints{F. Haberl, \email{fwh@mpe.mpg.de}\\
$^*$Based on observations obtained with XMM-Newton, an ESA science 
    mission with instruments and contributions directly funded by 
    ESA Member States and the USA (NASA)}
 
\institute{Max-Planck-Institut f\"ur extraterrestrische Physik,
               Giessenbachstra{\ss}e, 85748 Garching, Germany \and
           University of Western Sydney Nepean, P.O. Box 10,
               Kingswood, NSW 2747, Australia \and
           Australia Telescope National Facility, CSIRO, P.O. Box
               76, Epping, NSW 2121, Australia }
 
\date{Received 29 September 2000; Accepted 25 October 2000}
 
\abstract{
The XMM-Newton first-light image revealed X-ray point sources which
show heavily absorbed power-law spectra. The spectral indices and 
the probable identification of a radio counterpart for the brightest 
source suggest AGN shining through the interstellar gas of the 
Large Magellanic Cloud (LMC). The column densities derived from the X-ray 
spectra in combination with H\,{\sc i} measurements will allow to draw 
conclusions on H\,{\sc i} to H$_2$ ratios in the 
LMC and compare these with values found for the galactic plane.
}

\maketitle

\section{Introduction}

The study of active galactic nuclei and quasi stellar objects (both
named AGN hereafter) behind nearby galaxies allows to probe the
interstellar matter in the foreground galaxy. Absorption lines in
optical spectra provide information about the absorbing material along
the line of sight through the galaxy. However, very few AGN are known
behind nearby galaxies as surveys for these objects avoided the crowded
fields. Dedicated searches for AGN behind nearby galaxies were performed
by Tinney et al. (\cite{tcz97}) and  Crampton et al. (\cite{cgc97}). 
Conducting optical
identification programs of ROSAT X-ray sources, they found 15 background
objects in the direction of the Small Magellanic Cloud and 11
towards the LMC involving also clusters of
galaxies and individual distant galaxies. For all these background
objects red-shifts were determined covering the range of 0.02 -- 2.3. From
more general optical identification programs of X-ray sources in the
Magellanic Clouds by Cowley et al. (\cite{cch84,csm97}) and
Schmidtke et al. (\cite{scc99}) 
further AGN were found and X-ray classification work
by Haberl \& Pietsch (\cite{hp99}) and Haberl et al. (\cite{hfp00}) 
suggests candidates for
background objects based on their X-ray (and radio) properties. However,
as consequence of the strong attenuation of X-rays in the ROSAT 0.1 --
2.4 keV band by interstellar matter which hampers the X-ray detection of
AGN in the dense cores (where column densities of the order of \ohcm{22}
are expected which suppress the flux at 1 keV by a factor $\sim$10 and at 
0.4 keV by more than 10$^4$), 
the background objects were mainly found in the
outer regions of the Magellanic Clouds.

The high sensitivity and good spectral resolution of the EPIC 
instruments on board of XMM-Newton (Jansen et al. \cite{j01}) 
over a wide energy band from 0.1 keV up to 15 keV provides 
a unique tool to detect hard (intrinsically and/or highly absorbed) 
X-ray sources and allows
for the first time to separate very accurately the intrinsic X-ray 
source spectrum (determined from higher energies) 
and the photo-electric absorption which attenuates the spectrum 
at lower energies. 
The EPIC instruments are therefore ideally 
suited to investigate the absorbing matter in the dense cores of 
nearby galaxies by observing AGN behind the galaxies.
While radio observations in the 21 cm line infer the column density of 
H\,{\sc i} only (e.g. Arabadijs \& Bregman \cite{ab99}), 
X-ray absorption measurements 
provide a linear measure of the total mass along the line of sight.

In this letter we report on a sample of highly absorbed point sources detected
in the XMM-Newton first-light image obtained by the EPIC-PN camera
(Str\"uder et al. \cite{s01}) south-west 
of the 30 Doradus region in the LMC. Their X-ray spectra
strongly suggest AGN as origin of the X-rays.

\section{Data analysis and results}

\subsection{XMM-Newton first light}
 
\begin{figure}
\resizebox{\hsize}{!}{\includegraphics[clip,bb=275 450 360 555]{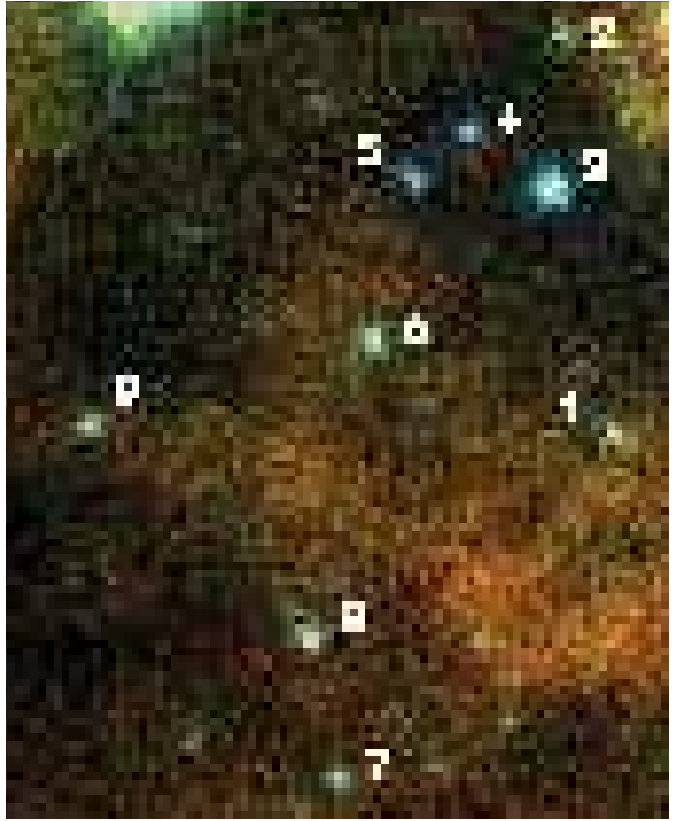}}
\caption{
  Section (7.7\arcmin\ x 9.5\arcmin) of the EPIC-PN first-light image 
  with the investigated point sources marked. The photon energy is coded 
  in colour from 0.3 keV (red) to 5.0 keV (blue). The bright source in the 
  upper-left is \object{N\,157B}. North is up and east to the left. For the 
  full image and further details on the production see Dennerl et al. 
  (\cite{dha01})}
\label{fig-xima}
\end{figure}

XMM-Newton received first light from a region of the LMC centered between 
the supernova remnants (SNRs) \object{N\,157B} and \object{SN\,1987A}. 
The field was 
observed five times with the EPIC-PN instrument in January 2000 using 
different optical light blocking filters and slightly different pointing 
positions (Tab.~\ref{tab-obs}).
The particular field in the LMC was chosen because of its richness
of X-ray emission seen already by the Einstein (Wang et al. 
\cite{whh91}) and ROSAT (Tr\"umper et al. \cite{tha91}) satellites. 
Besides diffuse emission from hot interstellar gas and emission from
extended sources like SNRs, the first-light image revealed point sources
which appear green and blue in the energy-colour coded image 
(Fig.~\ref{fig-xima}).
The colours indicate sources with hard X-ray spectrum as expected from
X-ray binaries residing in the LMC or background AGN shining through
the interstellar gas and dust of the LMC.

\begin{table}
\caption[]{XMM-Newton/EPIC-PN first-light observations of the LMC}
\begin{tabular}{rlc}
\hline\noalign{\smallskip}
Date~~~~~~~~~~    & Filter & Pointing Centre (J2000) \\
2000 January~~~~  &        & RA[h m s], Dec[\degr\ \arcmin\ \arcsec] \\
\noalign{\smallskip}\hline\noalign{\smallskip}
   19 16:19-17:22 & medium & 05 36 57, $-$69 13 47 \\
19/20 17:30-04:27 & medium & 05 36 57, $-$69 13 47 \\
   21 15:37-19:37 & medium & 05 37 04, $-$69 13 00 \\
21/22 20:32-07:00 & thin1  & 05 37 04, $-$69 13 00 \\
   22 09:07-12:01 & thin1  & 05 37 04, $-$69 13 00 \\
\noalign{\smallskip}\hline
\end{tabular}
\label{tab-obs}
\end{table}

\begin{table}
\caption[]{Hard X-ray point sources}
\begin{tabular}{cccccc}
\hline\noalign{\smallskip}
~   & RA (J2000) & Dec (J2000)               & CR$^1$ & HR       & Flux$^2$ \\
    & [h m s]      & [\degr\ \arcmin\ \arcsec] &        &          & \\
\noalign{\smallskip}\hline\noalign{\smallskip}
1 & 05 36 50.7  &  $-$69 16 40  &      &          &          \\ 
2 & 05 36 56.4  &  $-$69 11 49  &  3.8 & 0.45$\pm$0.23 & 1.6 \\ 
3 & 05 36 58.1  &  $-$69 13 43  & 25.1 & 1.16$\pm$0.06 &  19 \\ 
4 & 05 37 09.2  &  $-$69 13 00  &  3.2 & 1.60$\pm$0.22 & 3.4 \\ 
5 & 05 37 16.7  &  $-$69 13 33  &  3.9 & 3.00$\pm$0.24 & 4.4 \\ 
6 & 05 37 22.3  &  $-$69 15 31  &  5.2 & 0.55$\pm$0.15 & 3.1 \\ 
7 & 05 37 27.9  &  $-$69 20 47  &  3.7 & 0.57$\pm$0.68 & 1.9 \\ 
8 & 05 37 30.5  &  $-$69 19 04  &  3.6 & 0.35$\pm$0.19 & 2.1 \\ 
9 & 05 38 00.2  &  $-$69 16 28  &  3.7 & 0.66$\pm$0.25 & 1.4 \\ 
\noalign{\smallskip}\hline
\end{tabular}

$^1$ Count rate in $10^{-3}$ s$^{-1}$ in the 0.3 -- 10 keV band\\
$^2$ 0.3 -- 10 keV flux in 10$^{-14}$ erg cm$^{-2}$ s$^{-1}$
\label{tab-source}
\end{table}

First attempts of source detection using sliding window techniques
revealed at least a dozen of point sources. The brightest nine are 
summarized in Table~\ref{tab-source} and marked in Fig.~\ref{fig-xima}. 
The X-ray position was derived 
after alignment (translation only) of the image to match the coordinates 
of \object{SN\,1987A}. Systematic errors of up to 15\arcsec\ can be 
introduced by an uncertain rotation of the image of the order of 2\degr\
which is required to match the individual images. The coordinates
in Table~\ref{tab-source} should therefore be regarded as preliminary.
A hardness ratio to characterize the X-ray spectrum was obtained by 
dividing the counts in the 2.0 -- 10.0 keV and 0.5 -- 2.0 keV bands. 
As expected the blue sources in the colour image show highest hardness 
ratios above 1.0 while the green sources have values around 0.5.

\begin{figure}
\resizebox{8.3cm}{!}{\includegraphics[angle=-90,clip,bb=100 40 555 715]{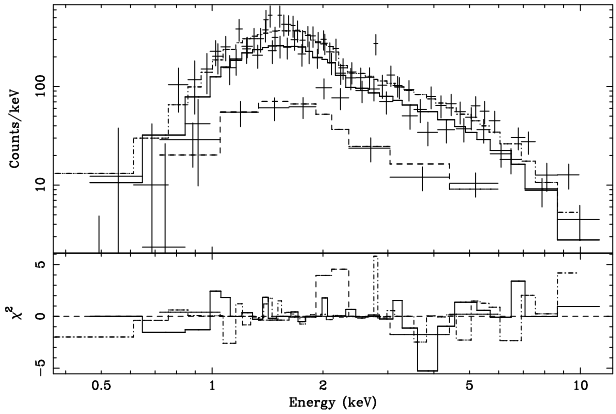}}
\caption{
  EPIC-PN spectra of \object{J\xmc} (source 3) obtained from three different 
  data sets together with the best fit model as histogram 
  (solid: merged observations 1 and 2, medium filter; 
  dash: observation 3, medium filter; 
  dash-dot: observations 4 and 5, thin filter). 
  The different normalizations reflect the different exposure 
  times}
\label{fig-s05}
\end{figure}

To derive X-ray spectra we merged the first two and the last two 
observations which had identical instrumental configurations. 
For each of the three resulting data sets a spectrum was 
derived from source and nearby background regions. Only for the brightest
source \object{J\xmc} (source 3), close to the center of the field of view,
spectra with sufficient statistics for a spectral fit were obtained.
The three spectra were simultaneously fit with a power-law model
attenuated by photo-electric absorption. A column 
density with equivalent \nh\ of 6\hcm{20} (Dickey et al. \cite{dl90}) 
with solar abundance was assumed
(fixed in the fit) for the galactic foreground absorption. For the LMC absorption
(free in the fit) a metallicity of 0.5 solar (Russell \& Dopita 
\cite{rd92}) was used.
Spectra and best-fit model (histogram) are shown in Fig.~\ref{fig-s05}.
Since no photon arrival times
were available in the first processing version of the data, no exposure
correction was applied to the spectra. 
The strong turn-over of the spectrum is caused by high photo-electric 
absorption.


\begin{figure}
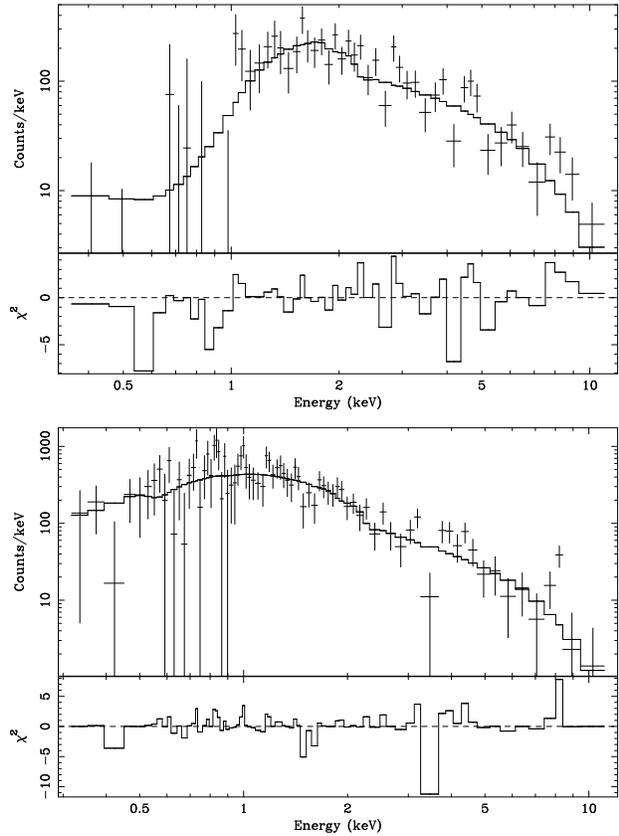

\resizebox{8.3cm}{!}{\includegraphics[angle=-90,clip,bb=100 40 555 715]{XMM08_f3a.eps}}
\resizebox{8.3cm}{!}{\includegraphics[angle=-90,clip,bb=100 40 555 715]{XMM08_f3b.eps}}
\caption{
  EPIC-PN spectra of point sources detected in the first light image: 
  combined \object{J\xmd} and \object{J\xme} (top) and 
  combined \object{J\xmf}, \object{J\xmi} and \object{J\xmh} (bottom)}
\label{fig-spectra}
\end{figure}

To increase photon statistics we merged spectra from sources with 
similar hardness ratios. Sources 4 and 5 are
located close to \object{J\xmc} (source 3) and show the highest hardness ratios.
We also merged the spectra of sources 6, 8 and 9 
with hardness ratios between 0.35 and 0.66. Again, highly absorbed 
power-law models were found to fit the spectra satisfactory (see 
Table~\ref{tab-fit} where \nh\ and photon index $\gamma$ are listed). 
For better comparison and presentation 
the three spectra from the individual data sets were merged into a 
single spectrum. These are shown for the combined sources 
in Fig.~\ref{fig-spectra}. The model is plotted using
an average response matrix. The combined spectrum of 
sources 6, 8 and 9 clearly
suffers less absorption compared to the spectrum of \object{J\xmc} (source 3)
as indicated by the hardness ratios and quantified by the \nh\ derived 
from the spectral fits.

To estimate the exposure times of the observations we produced EPIC-PN spectra 
of the SNR \object{N\,157B} and normalized their flux in the 0.3 -- 2.4 keV band to
a ROSAT PSPC spectrum of this source. Unfortunately during the first two
XMM observations \object{N\,157B} was located on a CCD border and flux was lost.
Therefore, we only used the third part of the observations (the two thin 
filter observations) for this purpose where we found a net exposure of 
31.0 ks. Source count rates and fluxes listed in Table~\ref{tab-source}
are derived from the merged thin filter observations using this 
exposure. The fluxes (0.3 -- 10 keV) were derived from the spectral fits
using the parameters given in Table~\ref{tab-fit}. For those sources
without spectral fit the parameters from the combined fit including
\object{J\xmf} were used, i.e. low absorption compatible with the
hardness ratios. 

\begin{table}
\caption[]{Spectral parameters for a power-law model}
\begin{tabular}{lccc}
\hline\noalign{\smallskip}
Source & \nh\                   & $\gamma$ & $\chi^2$/dof \\
       & [10$^{22}$ cm$^{-2}$]  &          &\\
\noalign{\smallskip}\hline\noalign{\smallskip}
3      & 1.69$\pm0.25$          & 1.97$\pm0.16$ & 80.7/85 \\
4/5    & 2.26$\pm0.75$          & 1.75$\pm0.32$ & 69.8/49 \\
6/8/9  & 0.35$\pm0.13$          & 1.78$\pm0.23$ & 93.8/77 \\
\noalign{\smallskip}\hline
\end{tabular}
\label{tab-fit}
\end{table}

\subsection{The radio source \object{ATCA\,J0536.9$-$6913} = 
\object{J\xmc}}

The region around \object{J\xmc} was observed as part of 
ATCA mosaic observations of the LMC with a baseline of 1500\,m 
at frequencies of 1420 and 2370\,MHz with angular 
resolutions of $\sim$45\arcsec\ and $\sim$30\arcsec.
Similar ATCA observations in `snap-shot 
mode' at 4800\,MHz were undertaken for specific regions 
including \object{J\xmc} (Filipovi\'c \& Staveley-Smith \cite{fs98}).
The baseline of these observations was 375\,m with 
resolutions of $\sim$30\arcsec\ and $\sim$15\arcsec, respectively. 
An additional observation of this region was made 
at 843\,MHz with the Molonglo Synthesis Telescope (MOST) (Mills 
\cite{m81}) as part of LMC survey and \object{SN\,1987A} monitoring projects at 
this frequency.

Radio-continuum emission from a point-like source is detected 
at \mbox{RA\,=\,05$^{\rm h}$\,36$^{\rm m}$\,56\fs62} and 
\mbox{Dec\,=\,$-$69$\degr$\,13$\arcmin$\,27\farcs7} (J2000),
compatible with the X-ray position given the uncertainty
in position angle of the X-ray image. 
The integrated flux density of this source, designated 
\object{ATCA\,J0536.9$-$6913}, at 843, 1420, 2370 and 4800\,MHz was determined
to 336$\pm$30, 211$\pm$20, 116$\pm$11 and 102$\pm$10~mJy, respectively.

\begin{figure}
\resizebox{\hsize}{!}{\includegraphics[angle=-90,clip,bb=210 40 575 430]{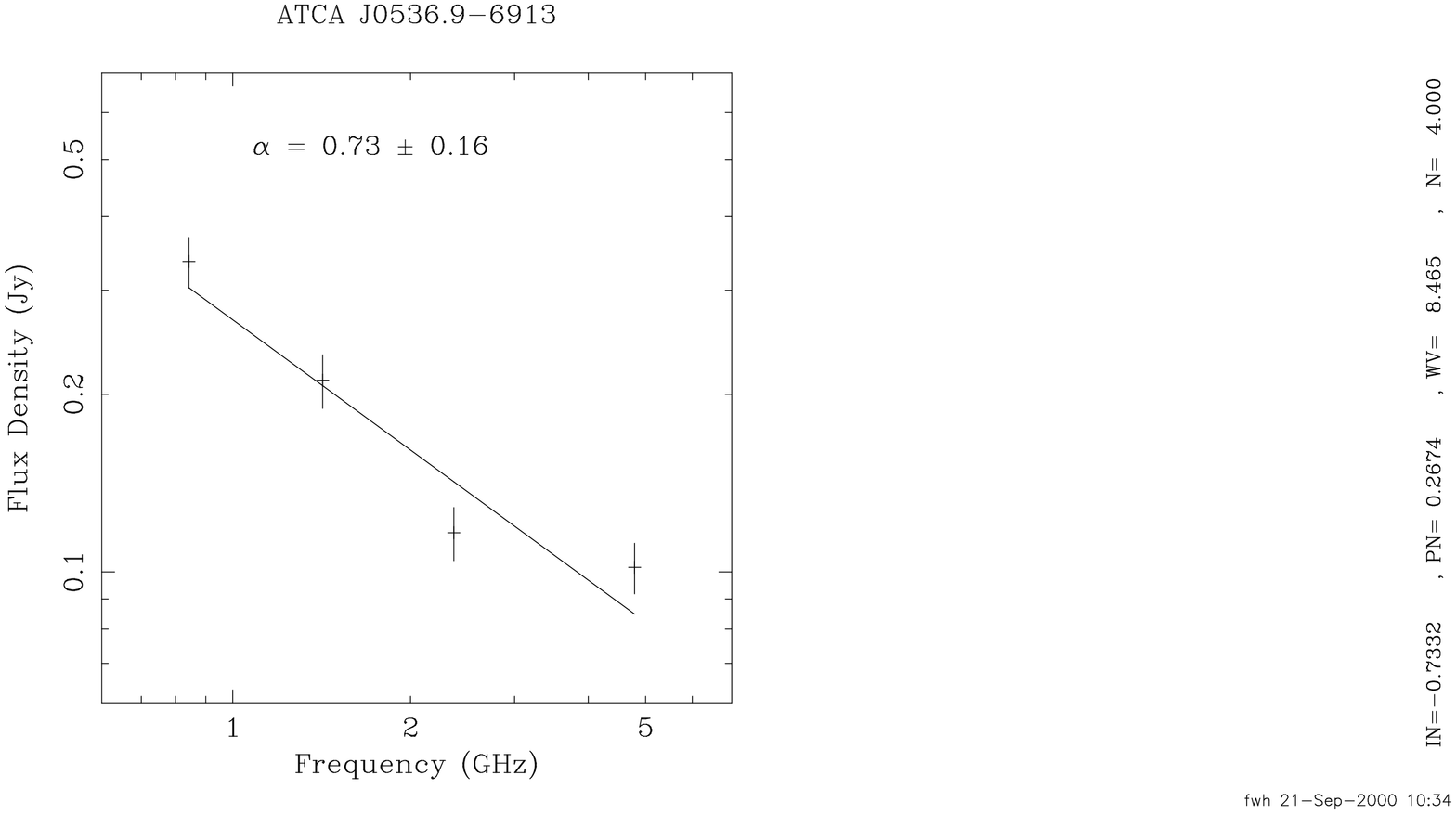}}
\caption[]{Radio spectrum of \object{ATCA\,J0536.9$-$6913} = \object{J\xmc}}
\label{fig-radio}
\end{figure}

We derive a spectral index $\alpha$ = 0.73$\pm$0.16 
(defined by \mbox{$\rm S_{\nu}\sim \nu^{-\alpha}$}) from the 
flux densities, $\rm S_{\nu}$, at 
the frequencies, $\nu$, of 843, 1420, 2370 and 4800\,MHz
(Fig.~\ref{fig-radio}).
This spectral index is typical for radio background sources which 
are mainly AGN or radio quasars (Filipovi\'c et al. \cite{fhw98}). 

\section{Discussion}

We investigated nine point sources with hard X-ray spectra detected
in the XMM-Newton first-light images. Their X-ray spectra can be 
represented by highly absorbed power-laws and the derived photon indices
between 1.7 and 2.0 are typical for steep AGN spectra (e.g. 
Vaughan et al. \cite{vrw99}).
In contrast accreting high mass X-ray binaries show harder X-ray spectra 
with typical indices between 0 and 1 (e.g. Yokogawa et al. \cite{yit00}) not 
compatible with the indices found for our point sources. 
The radio spectral index of 0.73 found for \object{J\xmc} 
also supports the AGN interpretation.

AGN spectra sometimes show intrinsic absorption which can 
easily reach the amount
seen in the spectra reported here. So the question remains which
fraction of the column density can be attributed to the LMC. The fact
that the three close point sources near the center of the field of view
all show similar high absorption of the order of \ohcm{22} suggests that
most of the \nh\ is intrinsic to the LMC. The derived column density
of 0.35\hcm{22} of sources only $\sim$2\arcmin\ away from the central
sources shows that the absorption can change by a factor of more than 5
over distances of $\sim$30 pc in the LMC.
It is remarkable in this respect that the north-west part of the first-light
image lacks of point sources. This suggests that even higher column densities 
exceeding \ohcm{22} are required to suppress the detection of AGN in 
that region. 

X-ray absorption measurements are directly related to the total
hydrogen column density, comprising H\,{\sc i}, H$_2$ and H\,{\sc ii}
column densities. Because the absorbing elements at X-ray energies are 
He and metals their relative abundance must be taken into account
to derive X-ray hydrogen column densities. This was done in our 
spectral fits assuming abundances of 0.5 relative to solar (Russell \& 
Dopita \cite{rd92}). Kim et al. (\cite{kss00}) have produced
a H\,{\sc i} column density map of the LMC with a resolution of
1\arcmin\ which shows variations from 0 to 5\hcm{21} with maximum values
around the 30\,Doradus region. The 
higher absorption seen in the XMM sources suggests 
either large amounts of H\,{\sc ii} and/or H$_2$ gas or higher metal abundances 
along their lines of sight through the LMC. Higher abundances of O, Ne, 
Mg and Si in the interstellar medium are suggested by 
the spectrum of the diffuse emission near the 30 Dor region 
(Dennerl et al. \cite{dha01}).

\begin{acknowledgements}
The XMM-Newton project is supported by the Bundesministerium f\"ur
Bildung und Forschung / Deutsches Zentrum f\"ur Luft- und Raumfahrt 
(BMBF/DLR), the Max-Planck Society and the Heidenhain-Stiftung.
We thank A. Green from the University of Sydney for providing the data 
of the MOST observations.
\end{acknowledgements}

\end{document}